\pdfoutput=1
\documentclass[english,journal]{IEEEtran}
\usepackage[latin9]{inputenc}
\usepackage[T1]{fontenc}
\usepackage{ae}
\usepackage{aecompl}
\usepackage{mathrsfs}
\usepackage{amsmath}
\usepackage{amssymb}
\usepackage{graphicx}

\makeatletter

\providecommand{\tabularnewline}{\\}

\@ifundefined{lettrine}{\usepackage{lettrine}}{}

\ifCLASSINFOpdf
\else
\fi

\usepackage{amsmath}

\makeatother

\begin{document}

\title{Directional Spatial Channel Estimation For Massive FD-MIMO in Next
Generation 5G Networks}

\author{\IEEEauthorblockN{Ali A. Esswie\textit{$^{\ast}$}, \textit{Student Member, IEEE}, Octavia
A. Dobre\textit{$^{\ast}$}, \textit{Senior} \textit{Member, IEEE},
and Salama Ikki\textit{$^{\ddagger}$}, \textit{Member, IEEE}\\
\textit{$^{\ast}$}Faculty of Engineering and Applied Science, Memorial
University, St. John\textquoteright s, Canada\textit{}\\
\textit{$^{\ddagger}$}Department of Electrical Engineering, Lakehead
University, Thunder Bay, Canada}}
\maketitle
\begin{abstract}
Full-dimensional (FD) channel state information at transmitter (CSIT)
has always been a major limitation of the spectral efficiency of cellular
multi-input multi-output (MIMO) networks. This letter proposes an
FD-directional spatial channel estimation algorithm for frequency
division duplex massive FD-MIMO systems. The proposed algorithm uses
the statistical spatial correlation between the uplink (UL) and downlink
(DL) channels of each user equipment. It spatially decomposes the
UL channel into azimuthal and elevation dimensions to estimate the
array principal receive responses. An FD spatial rotation matrix is
constructed to estimate the corresponding transmit responses of the
DL channel, in terms of the frequency band gap between the UL and
DL channels. The proposed algorithm shows significantly promising
performance, approaching the ideal perfect-CSIT case without UL feedback
overhead. 

\textit{Index Terms}\textemdash{} Full-dimensional MIMO; spatial correlation;
frequency division duplex; 5G; massive MIMO; CSI.
\end{abstract}


\thispagestyle{empty}

\IEEEpeerreviewmaketitle{}

\section{Introduction}

\IEEEPARstart{F}{}ull dimensional massive multi-input multi-output
(FD-mMIMO) is a key technology for boosting the spectral efficiency
(SE) of 5G cellular networks {[}1{]}. Performance improvement of FD-mMIMO
systems is achieved by using adaptive transmission at the base-station
(BS) over the FD cell space. However, the assumption of perfect FD
channel state information at the transmitter (FD-CSIT) is vital for
achieving optimality, which is not feasible in practice {[}2, 3{]}. 

Hence, typical CSIT acquisition algorithms in time division duplex
systems reasonably assume channel reciprocity, where the downlink
(DL) channel can be approximated by the transpose of the uplink (UL)
channel. In frequency division duplex (FDD) systems, channel reciprocity
is not applicable due to the frequency band offset $\Omega$. Consequently,
channel quantization and limited-feedback algorithms have been widely
considered {[}4{]}. Current LTE-A Pro standards {[}5{]} define double-codebooks
for tracking the channels small- and large-scale variations. For FD-mMIMO
systems, the double-codebooks are extended to scan the azimuthal and
elevation dimensions, providing a Kronecker-product (KP) based beamforming
algorithm {[}6{]}. 

However, channel quantization represents a major limitation of the
network spatial degrees of freedom (DoFs), regardless of the number
of transmit antennas {[}7{]}. Hence, the design of the beamformed
CSI-reference-signals (CSI-RS) is widely studied in recent standards
{[}8, 9{]}, where the DL pilots are distributed across several FD
beamforming directions. CSI-RS algorithms have shown scalability and
performance improvement with limited feedback overhead; however, they
may result in blind coverage spots if scanning precision is insufficient.
Furthermore, an FDD mMIMO system based on DL spatial channel estimation
(FMMSCE) {[}7{]} has been recently proposed, to remove the limitation
of the channel quantization; though, it suffers from sub-optimal performance
in FD systems due to the missing elevation DoFs. 

In this work, a directional spatial channel estimation (D-SCE) algorithm
is proposed for FDD FD-mMIMO systems. The UL channel is spatially
projected over the FD space of a pre-designed discrete Fourier transform
(DFT) codebook. The FD spatial power spectrum of the UL channel is
estimated to obtain the instantaneous receive response of the antenna
array. The estimated array response is spatially rotated in terms
of $\Omega$ to compensate for the spatial deviation of the principal
DL channel clusters and thereby attain the corresponding transmit
response. Finally, the UL channel is spatially beamformed towards
the principal set of the estimated transmit responses. The minimum
mean square error (MMSE) criterion is applied to refine the estimation
accuracy. The proposed D-SCE algorithm shows promising SE improvement,
without the requirement of channel quantization or feedback overhead. 

This paper is organized as follows. In Section II, the spatial channel
modeling is presented. Section III introduces the proposed D-SCE algorithm.
Performance results are discussed in Section IV and conclusions are
drawn in Section V.

\textit{Notations:} $(x)^{\textnormal{T}}$ , $(x)^{\textnormal{H}}$
and $(x)^{\textnormal{-1}}$ denote the transpose, Hermitian, and
inverse operations on $x$. $x\,\text{\ensuremath{\otimes}}\,y$ stands
for the Kronecker product of $x$ and $y$, while $\overline{x}$
and $|x|$ represent the mean and absolute value of $x.$ $\ensuremath{x\sim\mathbb{C\ensuremath{\mathbb{N}}}}(0,\sigma^{2})$
denotes a complex Gaussian random variable with zero mean and variance
$\sigma^{2}$, while $\left\{ x\right\} $ indicates the set of possible
$x$ values. $x_{\kappa},\kappa\text{\ensuremath{\in}}\{\textnormal{ul},\textnormal{dl}\}$
denotes the link direction of $x$. 

\section{System Model}

In this work, we consider a DL multi-user FD mMIMO system, as shown
in Fig. 1. A BS is mounted with $N_{t}=N_{v}\times N_{h}$ planar
uniform rectangular array (URA), where $N_{v}$ and $N_{h}$ are the
elevation and azimuthal antenna elements, respectively. There are
$K$ FD uniformly distributed users with $M_{r}$ antennas. Each UE
accepts $d$ independent DL spatial streams $x\sim\text{\ensuremath{\mathbb{C\ensuremath{\mathbb{N}}}}(0,\ensuremath{\frac{P}{d}\boldsymbol{\mathrm{I}}_{d}}), }$where
$P$ is the user transmit power. The received DL signal at the $k^{th}$
user is expressed as

\begin{equation}
\boldsymbol{y}_{k}=\boldsymbol{\mathrm{\boldsymbol{\textnormal{\textbf{H}}}}}_{\textnormal{dl},k}\boldsymbol{\textnormal{\textbf{V}}}_{k}\boldsymbol{\mathrm{\boldsymbol{\textnormal{\textbf{x}}}}}_{k}+\sum_{j=0,j\neq k}^{K-1}\boldsymbol{\mathit{\textnormal{\textbf{H}}}}_{\textnormal{dl,\textit{k}}}\boldsymbol{\textnormal{\textbf{V}}}_{j}\boldsymbol{\textnormal{\textbf{x}}}_{j}+\boldsymbol{\boldsymbol{\mathrm{\boldsymbol{\textnormal{\textbf{n}}}}}}_{\boldsymbol{k}},
\end{equation}
where $\boldsymbol{\mathrm{H}}_{\textnormal{\textnormal{dl},\textit{k}}}\in\ensuremath{\mathbb{C}}^{M_{r}\times N_{t}},\forall k\in\{0,1,\ldots,K-1\}$
is the DL spatial channel of the $k^{th}$ user, $\boldsymbol{\mathrm{\boldsymbol{\textnormal{\textbf{V}}}}}_{k}\in\ensuremath{\mathbb{C}}^{N_{t}\times1}$
is the zero-forcing precoding matrix given as $\boldsymbol{\textnormal{\textbf{V}}}_{k}=\boldsymbol{\left(\boldsymbol{\mathrm{H}}_{\textnormal{\textnormal{dl},\textit{k}}}\right)}^{\textnormal{H}}\left(\boldsymbol{\boldsymbol{\mathrm{H}}_{\textnormal{\textnormal{dl},\textit{k}}}}\left(\boldsymbol{\mathrm{H}}_{\textnormal{\textnormal{dl},\textit{k}}}\right)^{\textnormal{H}}\right)^{-1},$
and $\boldsymbol{\boldsymbol{\mathrm{\boldsymbol{\textnormal{\textbf{n}}}}}}_{\boldsymbol{k}}$
is the additive white Gaussian noise. We adopt the spatially-correlated
channel model {[}10{]}, where the channel is described by its major
$C$ scattering clusters, spatially distributed over the FD cell space
with $Z$ rays per cluster. The DL channel coefficient from the $n^{th}$
transmit antenna to the $m^{th}$ receive antenna is given by 
\begin{figure}
\begin{centering}
\includegraphics[scale=0.3]{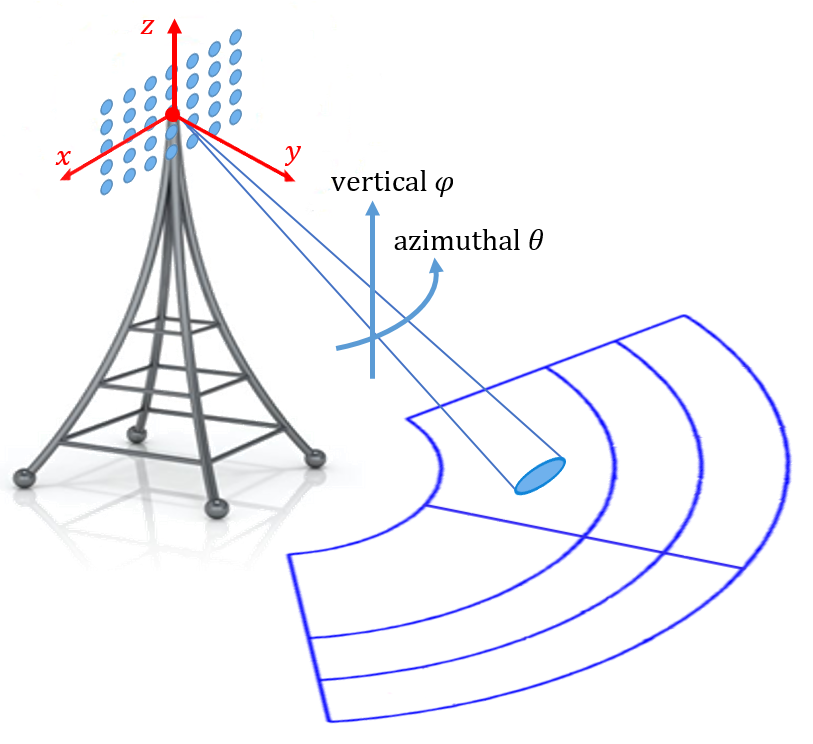}
\par\end{centering}
\centering{}\caption{Coverage footprint of the FD mMIMO system modeling.}
\end{figure}

\begin{equation}
h_{(m,n)_{k}}=\frac{1}{\sqrt{C}}\sum_{c=0}^{C-1}\sqrt{\alpha_{k}}\text{\,\ensuremath{\mathcal{G}}}_{c,k}\,r_{(m,n,c)_{k}},
\end{equation}
where $\alpha_{k}=\ell\epsilon_{k}^{\beta}\mu_{k}$ is the channel
large-scale factor, $\ell$ is a propagation constant, $\mu_{k}$
is the shadow fading factor, and $\epsilon_{k}^{\beta}$ is the separation
distance, with $\beta$ as the pathloss exponent, and $\text{\ensuremath{\mathcal{G}}}_{c,k}\sim\text{\ensuremath{\mathbb{C\ensuremath{\mathbb{N}}}}(0,1). }$The
steering element $r_{(m,n,c)_{k}}$ of the channel coefficient is
given by 

$r_{(m,n,c)_{k}}=$

\begin{equation}
\sqrt{\frac{\xi\psi}{Z}}\sum_{z=0}^{Z-1}\left(\begin{array}{c}
\sqrt{\text{\ensuremath{\mathfrak{D}}}_{\textnormal{BS}}^{m,n,c,z}(\theta_{\textnormal{AoD}},\varphi_{\textnormal{EoD}})}\,e^{j\textnormal{(\text{\ensuremath{\eta}}\textit{d}\ensuremath{\overline{f}}+\ensuremath{\Phi_{m,n,c,z}})}}\\
\times\sqrt{\text{\ensuremath{\mathfrak{D}}}_{\textnormal{UE}}^{m,n,c,z}(\theta_{\textnormal{AoA}},\varphi_{\textnormal{EoA}})}\,e^{j(\text{\ensuremath{\eta}}d\sin(\theta_{\textnormal{m,n,c,z,AoA}}))}\\
\times e^{j\text{\ensuremath{\eta}}||s||\cos(\varphi_{\textnormal{m,n,c,z,EoA}})\cos(\theta_{\textnormal{m,n,c,z,AoA}}-\theta_{s})t}
\end{array}\right),
\end{equation}
where $\xi$ and $\psi$ are the power and large-scale coefficient,
$\text{\ensuremath{\mathfrak{D}}}_{\textnormal{BS}}^{m,n,c,z}$ and
$\text{\ensuremath{\mathfrak{D}}}_{\textnormal{UE}}^{m,n,c,z}$ are
the BS and UE spatial antenna patterns, $\text{\ensuremath{\eta}}$
is the wave number, $\theta$ denotes the azimuthal angle of arrival
$\theta_{\textnormal{AoA}}$ and departure $\theta_{\textnormal{AoD}}$,
while $\varphi$ denotes the elevation angle of arrival $\varphi_{\textnormal{EoA}}$
and departure $\varphi_{\textnormal{EoD}}$, respectively. $s$ is
the speed of the $k^{th}$ user, $\overline{f}=f_{x}\cos\theta_{\textnormal{AoD}}$
$\cos\varphi_{\textnormal{EoD}}+f_{y}\cos\varphi_{\textnormal{EoD}}\sin\theta_{\textnormal{AoD}}+f_{z}\sin\varphi_{\textnormal{EoD}}$
is the generic displacement vector of the transmit antenna array. 

\section{Proposed Directional Channel Estimation for FD-mMIMO Networks}

\subsection{Spatially-Correlated FD-MIMO Channels}

The exact spatially-correlated FD-MIMO channel model, introduced in
Section II, can be rewritten only by its predominant spatial clusters,
distributed over the FD cell space as

\begin{equation}
\boldsymbol{\textnormal{\textbf{H}}}_{\kappa}=\frac{1}{\sqrt{C}}\sum_{c=0}^{C-1}g_{\kappa,c}\textnormal{\textbf{a}}_{\kappa,c}(\phi_{c}),
\end{equation}
where $\boldsymbol{\textnormal{\textbf{H}}}_{\kappa},\kappa\text{\ensuremath{\in}}\{\textnormal{ul},\textnormal{dl}\}$
is the UL/DL spatial channel matrix of an arbitrary user, $g_{\kappa,c}$
is the $c^{th}$ cluster gain of the UL/DL channel, and $\textnormal{\textbf{a}}_{\kappa,c}(\phi_{c})$
is the UL receive or DL transmit antenna FD response of the $c^{th}$
cluster, with $\phi_{c}$ as the FD spatial angle of the corresponding
antenna response. The FD antenna response $\textnormal{\textbf{a}}_{\kappa,c}(\phi_{c})$
is composed of the horizontal and vertical responses by the Kronecker
product as $\textnormal{\textbf{a}}_{\kappa,c}(\phi_{c})=\textnormal{\textbf{a}}_{\kappa,c}^{h}(\theta_{c})\text{ \ensuremath{\otimes} }\textnormal{\textbf{a}}_{\kappa,c}^{v}(\varphi_{c})$,
where the horizontal $\textnormal{\textbf{a}}_{\kappa,c}^{h}(\theta_{c})$
and vertical $\textnormal{\textbf{a}}_{\kappa,c}^{v}(\varphi_{c})$
antenna response, in the azimuthal direction $\theta_{c}$ and elevation
direction $\varphi_{c}$ of the $c^{th}$ cluster, are given by

\begin{equation}
\textnormal{\textbf{a}}_{\kappa,c}^{h}(\theta_{c})=\left[1,e^{-j2\pi\Delta_{\kappa}^{h}\cos\theta_{c}},\ldots,e^{-j2\pi\Delta_{\kappa}^{h}(N_{h}-1)\cos\theta_{c}}\right]^{\textnormal{T}},
\end{equation}

\begin{equation}
\textnormal{\textbf{a}}_{\kappa,c}^{v}(\varphi_{c})=\left[1,e^{-j2\pi\Delta_{\kappa}^{v}\cos\varphi_{c}},\ldots,e^{-j2\pi\Delta_{\kappa}^{v}(N_{v}-1)\cos\varphi_{c}}\right]^{\textnormal{T}},
\end{equation}
where $\Delta_{\kappa}^{h}$ and $\Delta_{\kappa}^{v}$ are the horizontal
and vertical antenna physical spacing, respectively. The $c^{th}$
FD cluster can be sampled in the DFT domain as

\begin{equation}
\mathscr{H}_{\kappa,c}(b)=\sum_{n=0}^{N_{t}-1}g_{\kappa,c}e^{-j2\pi\Delta_{\kappa}n\cos(\phi_{c})}e^{\frac{-j2\pi bn}{N_{t}}},\,b=0,\ldots,N_{t}-1,
\end{equation}
 where $\Delta_{\kappa}$ denotes the effective antenna spacing of
the entire antenna array. The magnitude of $\text{\ensuremath{\mathscr{H}}}_{\kappa,c}(b)$
is described by 

{\small{}
\begin{equation}
|\text{\ensuremath{\mathscr{H}}}_{\kappa,c}(b)|=\left|g_{\kappa,c}\right|\times\left|\frac{\sin\left(\frac{N_{t}}{2}\left(-2\pi\Delta_{\kappa}\sin\left(90-\phi_{c}\right)+\frac{2\pi}{N_{t}}b\right)\right)}{\sin\left(\frac{1}{2}\left(-2\pi\Delta_{\kappa}\sin\left(90-\phi_{c}\right)+\frac{2\pi}{N_{t}}b\right)\right)}\right|.
\end{equation}
}{\small \par}

From (8), the leakage of each channel cluster becomes range-limited
with the number of the transmit antennas. Hence, with large antenna
arrays, the channel dimension reduces to fewer and more predominant
clusters. This leads to significant estimation precision if the directions
of only the most predominant DL clusters are sufficiently approached.
Furthermore, the $c^{th}$ channel cluster gain is expressed as 

\begin{equation}
g_{\kappa,c}=\varLambda_{\kappa,c}\gamma_{\kappa,c}\Upsilon_{\kappa,c},
\end{equation}
where $\varLambda_{\kappa,c}$ is a constant to represent the transmit
power, and antenna gain. $\gamma_{\kappa,c}$ and $\Upsilon_{\kappa,c}$
are the large- and small-scale factors of the $c^{th}$ channel cluster.
In dense environments, where low mobility conditions are applicable,
e.g., 30 km/h, the surrounding scatterers and Doppler shift slowly
vary between two successive UL and DL transmissions of the same user.
Thus, it is reasonable to assume that the average gain of both channel
clusters is constant {[}10{]}, being given as 

\begin{equation}
\text{\ensuremath{\zeta}}=\frac{1}{N_{t}\sqrt{C}}\text{\ensuremath{\mathbb{E}\left(\sum_{c=0}^{C-1}g_{\kappa,c}\right)} },
\end{equation}
where $\text{\ensuremath{\mathbb{E}}}$ denotes the statistical expectation.
The optimal DL transmit response to maximize the received power of
the $c^{th}$ channel cluster is given by 
\begin{figure}
\begin{centering}
\includegraphics[scale=0.7]{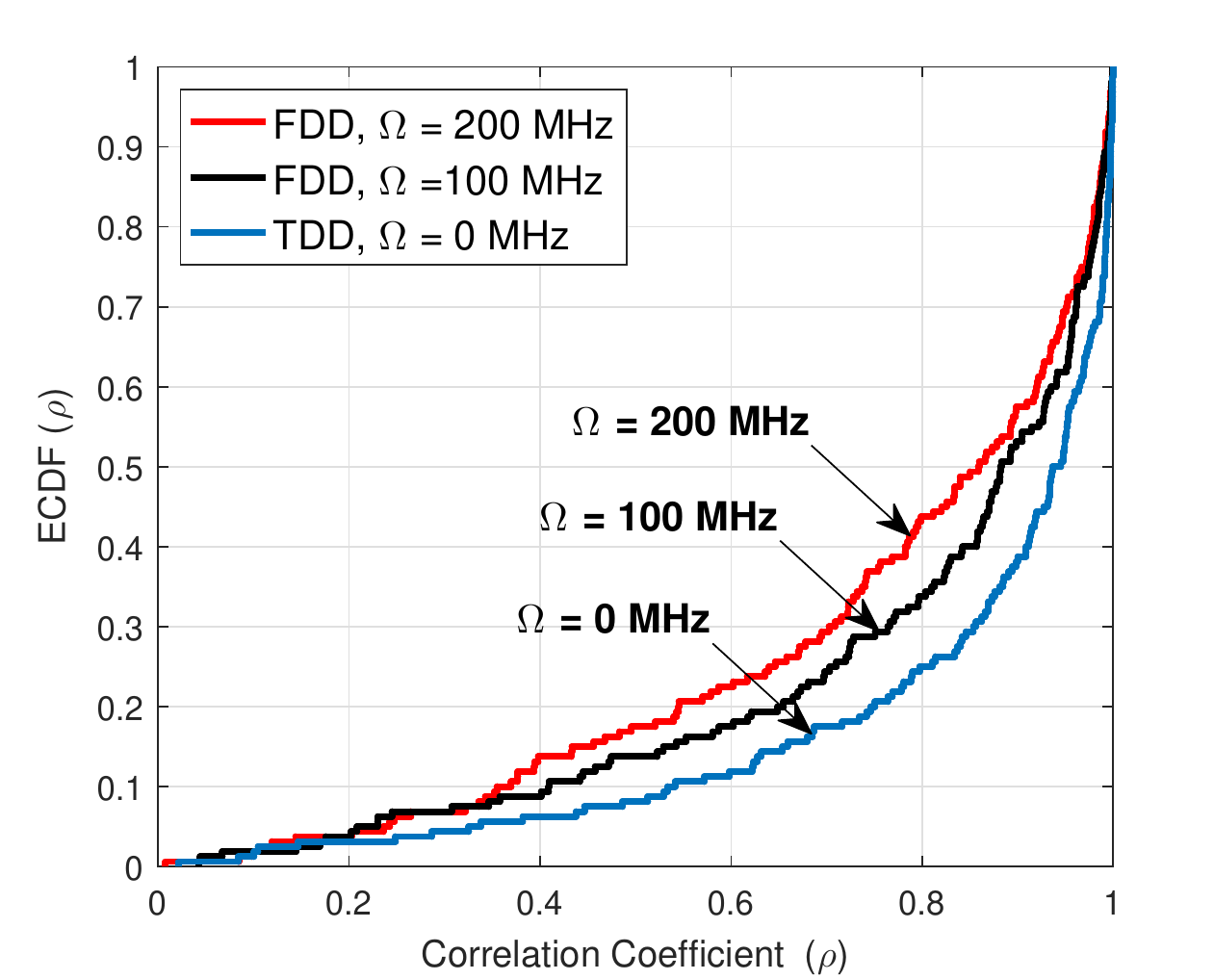}
\par\end{centering}
\centering{}\caption{ECDF ($\rho$).}
\end{figure}
\begin{figure}
\begin{centering}
\includegraphics[scale=0.49]{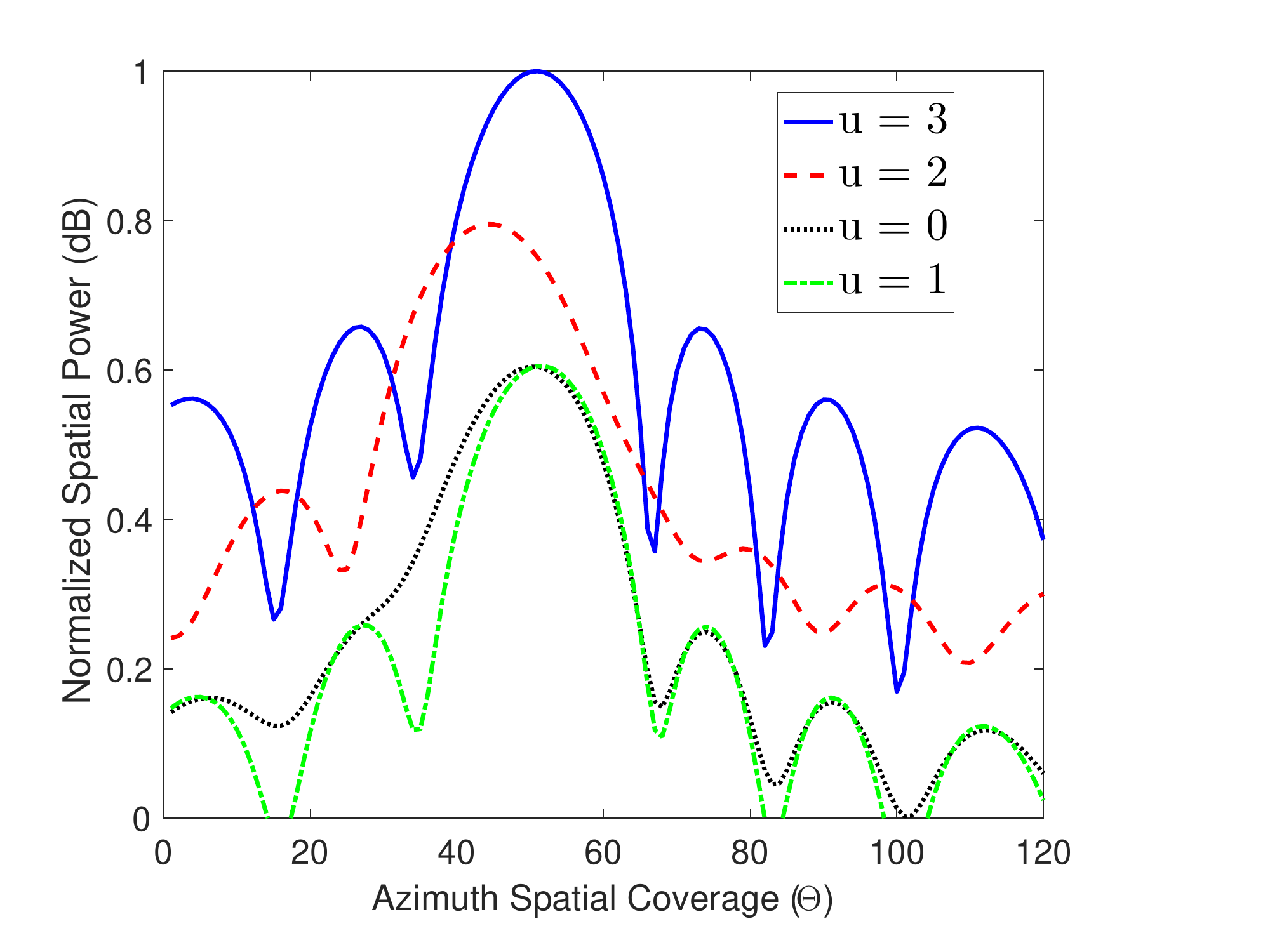}
\par\end{centering}
\centering{}\caption{FD UL spatial spectra.}
\end{figure}

\begin{equation}
\boldsymbol{\textnormal{\textbf{�}}}_{\textnormal{dl,\textit{c}}}(\hat{\phi_{c}})=\arg\max_{\phi}\left(\boldsymbol{\textnormal{\textbf{�}}}_{\textnormal{dl},c}^{\textnormal{H}}(\phi)\,\textnormal{\textbf{a}}_{\textnormal{dl,\textit{c}}}(\phi_{c})\,\textnormal{\textbf{a}}_{\textnormal{dl},c}^{\textnormal{H}}(\phi_{c})\,\boldsymbol{\textnormal{\textbf{�}}}_{\textnormal{dl},c}(\phi)\right),
\end{equation}
where $\boldsymbol{\textnormal{\textbf{�}}}_{\textnormal{dl,\textit{c}}}(\phi)$
and $\textnormal{\textbf{a}}_{\textnormal{dl,\textit{c}}}(\phi_{c})$
are the estimated and actual DL transmit responses at the BS. Thus,
the optimal transmit response $\boldsymbol{\textnormal{\textbf{�}}}_{\textnormal{dl,\textit{c}}}(\hat{\phi_{c}})$
should spatially align with the global set of the principal eigenvectors
of the actual DL response $\textnormal{\textbf{a}}_{\textnormal{dl,\textit{c}}}(\phi_{c}),$
where $\hat{\phi_{c}}=\phi_{c}.$ However, in FDD networks, the BS
only accesses the UL receive response $\textnormal{\textbf{a}}_{\textnormal{ul,\textit{c}}}(\phi_{c})$
since the transmit $\textnormal{\textbf{a}}_{\textnormal{dl,\textit{c}}}(\phi_{c})$
and receive $\textnormal{\textbf{a}}_{\textnormal{ul,\textit{c}}}(\phi_{c})$
antenna responses are not reciprocal, due to $\Omega$. Furthermore,
no closed-form relation between $\textnormal{\textbf{a}}_{\textnormal{dl,\textit{c}}}(\phi_{c})$
and $\textnormal{\textbf{a}}_{\textnormal{ul,\textit{c}}}(\phi_{c})$
exists in the literature, because it is a non-convex problem {[}5{]}.
Hence, the knowledge of the actual DL transmit response $\textnormal{\textbf{a}}_{\textnormal{dl,\textit{c}}}(\phi_{c})$
is not available at the BS. 

In this work, we transform the optimization problem in (11), with
the pre-knowledge requirement of the DL antenna response $\textnormal{\textbf{a}}_{\textnormal{dl,\textit{c}}}(\phi_{c})$,
into a search problem of the closest possible spatial directions,
observed from the spatial power spectrum of the UL channel, as it
will be discussed subsequently. 

Assuming a standard antenna sector of $120^{o}/90^{o}$ coverage in
both the azimuthal and elevation directions, the FD cell space is
spatially divided into $U$ and $Q$ elevation and azimuthal subspaces,
with an arbitrary scanning precision. Then, we define an FD-DFT beamforming
codebook at the BS to project the UL channel clusters over the virtual
beamforming sub-spaces. An approximate estimate of the spatial power
spectra of the UL/ DL channels, averaged over the $C$ channel clusters
within the entire FD space $\phi$ is given by

\begin{equation}
\textnormal{\textbf{P}}_{\kappa}(\phi)=\left[a_{\kappa}^{\textnormal{H}}(\phi)\left(\boldsymbol{\textnormal{\textbf{H}}}_{\kappa}\boldsymbol{\textnormal{\textbf{H}}}_{\kappa}^{\textnormal{H}}\right)^{-1}a_{\kappa}(\phi)\right]^{-1}.
\end{equation}
 Hence, the correlation coefficient of the UL and DL clusters is calculated
as 

\begin{equation}
\rho=\frac{\int\left(\textnormal{\textbf{P}}_{\textnormal{ul}}(\phi)-\overline{\textnormal{\textbf{P}}_{\textnormal{ul}}(\phi)}\right)\left(\textnormal{\textbf{P}}_{\textnormal{dl}}(\phi)-\overline{\textnormal{\textbf{P}}_{\textnormal{dl}}(\phi)}\right)d\phi}{\sqrt{\left(\textnormal{\textbf{P}}_{\textnormal{ul}}(\phi)-\overline{\textnormal{\textbf{P}}_{\textnormal{ul}}(\phi)}\right)^{2}}\sqrt{\left(\textnormal{\textbf{P}}_{\textnormal{dl}}(\phi)-\overline{\textnormal{\textbf{P}}_{\textnormal{dl}}(\phi)}\right)^{2}}}.
\end{equation}
The empirical cumulative density function (ECDF) of the correlation
coefficient is shown in Fig. 2, for different $\Omega$ values. As
can be noticed, the UL and DL spatial spectra, and hence, the receive
and transmit responses are highly correlated in the spatial domain,
due to the small channel spatial variance over the closely-spaced
antenna elements, e.g., for 50\% of the channel samples, a correlation
coefficient $\rho=0.8588$ is observed for $\Omega=200$ MHz.

Fig. 3 depicts the decomposable spatial spectra of the UL channel
across the entire azimuthal space of each elevation subspace, with
$U=4$ and $Q=120$. It is evident that the fourth elevation subspace
($u=3$) is the best-match-subspace for maximizing the received signal-to-interference-noise
ratio (SINR), capturing the additional UL spatial elevation DoFs.
It is worth mentioning that $Q\geqslant N_{t}$ should be satisfied
in order to fully utilize the spatial DoFs of the antenna array. Fig.
4 exemplifies the impact of $\Omega$ on the spatial deviation between
the observed UL AoAs and actual DL AoDs across the principal elevation
space\textit{.} It is worth noting that a spatial shift between the
most dominant clusters of the UL and DL channels appears, which is
a function of $\Omega$.

\subsection{FD Directional Spatial Channel Estimation }

The proposed D-SCE algorithm decomposes the observed UL channel into
2D projections over the azimuthal and elevation dimensions, using
a pre-designed FD beamforming codebook. The principal sub-array receive
responses in both dimensions are estimated to satisfy the maximization
of the spatial SINR. Then, the receive responses are spatially rotated,
in terms of $\Omega$, to estimate the BS transmit responses. The
observed UL channel is spatially beamformed towards the directions
of the estimated transmit responses. Finally, the MMSE criterion is
applied to enhance the estimation precision. 
\begin{figure}
\begin{centering}
\includegraphics[scale=0.7]{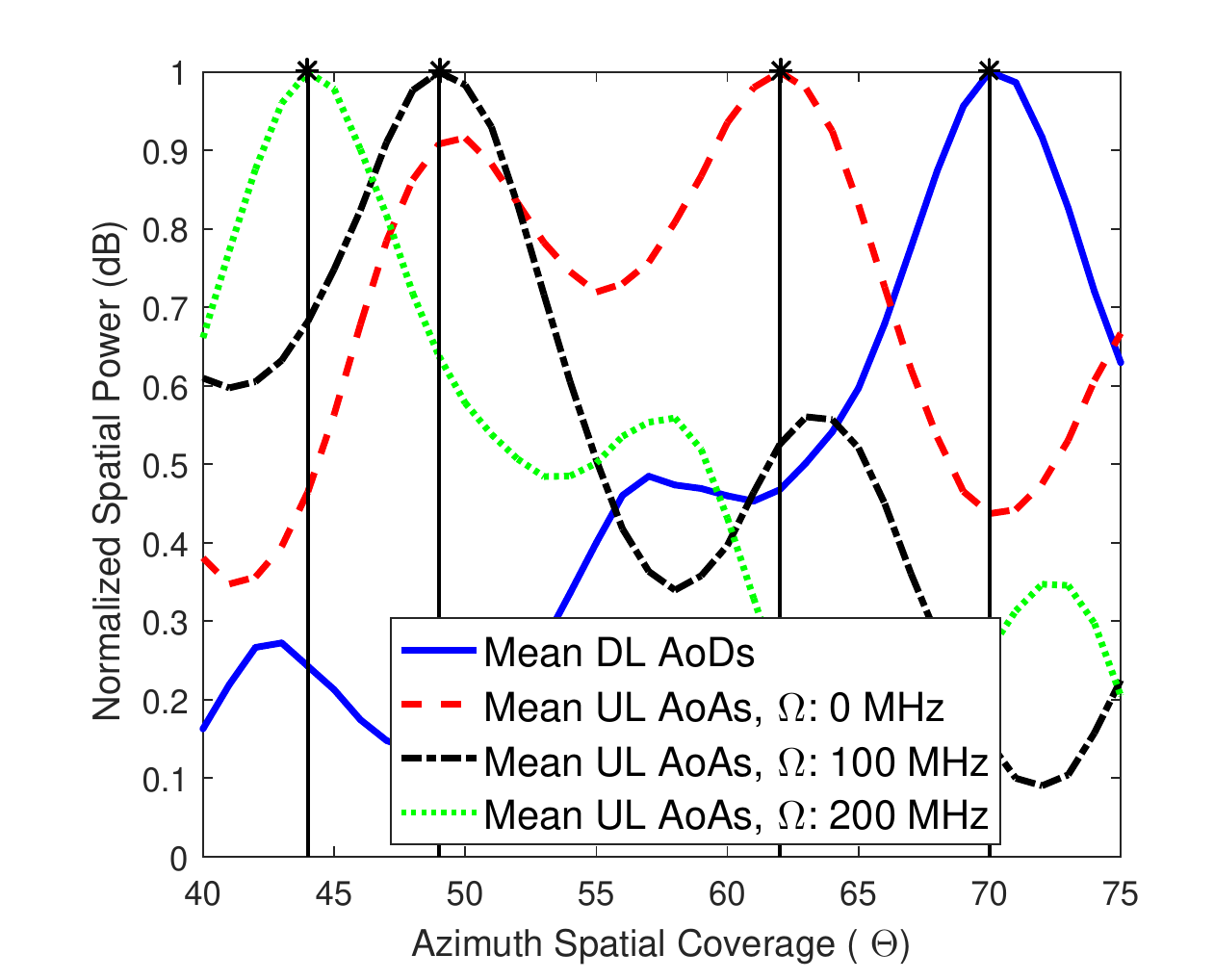}
\par\end{centering}
\centering{}\caption{Spatial deviation of the first principal UL/ DL cluster, with $\Omega$.}
\end{figure}

First, an arbitrary FD beamforming codeword $\boldsymbol{\textnormal{\textbf{W}}}_{\kappa}(\theta,\varphi)\in\ensuremath{\mathbb{C}}^{N_{t}\times1}$
is composed as

\begin{equation}
\boldsymbol{\textnormal{\textbf{W}}}_{\kappa}(\theta,\varphi)=\frac{1}{\sqrt{N_{t}}}\left(\textnormal{\textbf{a}}_{\kappa}^{h}(\theta)\text{\ensuremath{\,\otimes}}\,\textnormal{\textbf{a}}_{\kappa}^{v}(\varphi)\right).
\end{equation}
Accordingly, an FD beamforming spatial codebook is constructed as
$\left\{ \boldsymbol{\textnormal{\textbf{W}}}_{\kappa}(\theta_{q},\varphi_{u})\right\} $,
to cover the FD antenna sector by the discrete direction set $\left\{ \theta_{q},\varphi_{u}\right\} ,\forall q=0,1,\ldots Q-1,u=0,1,\ldots,U-1$.
Next, the UL FD spatial spectrum is estimated according to (12), where
the array responses are substituted by the FD codewords from the codebook
as $a_{\kappa}(\phi)=\boldsymbol{\textnormal{\textbf{W}}}_{\kappa}(\theta_{q},\varphi_{u})$.
The principal elevation subspace $\hat{\varphi_{u}}$ of the UL spatial
spectrum is obtained based on the maximization of the average received
power over the corresponding \textit{Q-codeword} azimuthal discrete
space $\left\{ \theta_{q}\right\} $ as 

\begin{equation}
\hat{\varphi_{u}}=\arg\underset{\left\{ \varphi_{u}\right\} }{\max}\left(\mathbb{E}\left(\left\Vert \boldsymbol{\textnormal{\textbf{P}}}_{\textnormal{ul}}(\theta_{q},\varphi_{u})\right\Vert ^{2}\right)\right).
\end{equation}

The strongest $N_{t}$ azimuthal directions are estimated as

\begin{equation}
\hat{\theta}_{n}=\arg\max_{\left\{ \theta_{q}\right\} }\left(\left\Vert \boldsymbol{\textnormal{\textbf{P}}}_{\textnormal{ul}}(\theta_{q},\hat{\varphi_{u}})\right\Vert ^{2}\right),\forall n=0,1,\ldots,N_{t}-1.
\end{equation}

The FD array principal receive response matrix $\boldsymbol{\textnormal{\textbf{A}}}_{\textnormal{R}}(\mathit{\Theta},\hat{\varphi_{u}})\in\ensuremath{\mathbb{C}}^{N_{t}\times N_{t}}$
can be given by

{\small{}
\begin{equation}
\boldsymbol{\textnormal{\textbf{A}}}_{\textnormal{R}}(\mathit{\Theta},\hat{\varphi_{u}})=\left[\left(\textnormal{\textbf{a}}_{\textnormal{ul}}^{h}(\hat{\theta}_{0})\otimes\textnormal{\textbf{a}}_{\textnormal{ul}}^{v}(\hat{\varphi_{u}})\right)^{\textnormal{T}}\ldots\left(\textnormal{\textbf{a}}_{\textnormal{ul}}^{h}(\hat{\theta}_{N_{t}-1})\otimes\textnormal{\textbf{a}}_{\textnormal{ul}}^{v}(\hat{\varphi_{u}})\right)^{\textnormal{T}}\right],
\end{equation}
}where $\mathit{\Theta}=\left\{ \hat{\theta}_{n}\right\} _{n=0}^{N_{t}-1}$
is the discrete set of the estimated principal azimuthal subspaces.
A spatial rotation matrix $\mathbf{\Gamma}\in\ensuremath{\mathbb{C}}^{N_{t}\times N_{t}}$
is constructed to compensate for the frequency band gap between the
UL and DL channels, with each rotation column vector given by

\begin{equation}
\text{\textbf{\ensuremath{\text{\textbf{\ensuremath{\mathit{\boldsymbol{\mathit{\digamma}}}}}}}=\ensuremath{\left[1,e^{-j2\pi\left(\frac{f_{dl}}{f_{ul}}\right)},\ldots,e^{-j2\pi\left(N_{t}-1\right)\left(\frac{f_{dl}}{f_{ul}}\right)}\right]}}}^{\textnormal{T}},
\end{equation}
where $f_{dl}$ and $f_{ul}$ are the operating center frequencies
of the DL and UL channels, respectively. The corresponding FD transmit
response matrix $\boldsymbol{\textnormal{\textbf{A}}}_{\textnormal{T}}(\varPhi,\mathit{\Psi})\in\ensuremath{\mathbb{C}}^{N_{t}\times N_{t}}$
is estimated by spatially rotating the obtained receive responses
through $\text{\ensuremath{\mathbf{\Gamma}}}$ by 

\begin{equation}
\boldsymbol{\textnormal{\textbf{A}}}_{\textnormal{T}}(\varPhi,\mathit{\Psi})=\boldsymbol{\textnormal{\textbf{A}}}_{\textnormal{R}}(\mathit{\Theta},\hat{\varphi_{u}})\textnormal{\ensuremath{\mathbf{\Gamma}^{\textnormal{T}}},}
\end{equation}
where $\varPhi$ and $\mathit{\Psi}$ denote a rotated set of the
azimuthal subspaces $\mathit{\Theta}$ and the elevation $\hat{\varphi_{u}}$
subspace. A rough estimate of the DL channel is calculated by beamforming
the observed UL channel over the estimated transmit responses as given
by

\begin{equation}
\boldsymbol{\textnormal{\textbf{H}}}_{\textnormal{dl}}^{(1)}=\boldsymbol{\textnormal{\textbf{H}}}_{\textnormal{ul}}^{\textnormal{H}}\boldsymbol{\textnormal{\textbf{A}}}_{\textnormal{T}}(\varPhi,\mathit{\Psi}).
\end{equation}

Finally, the estimated transmit response matrix is refined by applying
the MMSE criterion. The MMSE approach seeks a matrix $\textnormal{\textbf{G}}$
to minimize the corresponding MSE as 

\begin{equation}
\text{MSE = \ensuremath{\mathbb{E}}}\left\{ (\boldsymbol{\textnormal{\textbf{G}}}\boldsymbol{\textnormal{\textbf{H}}}_{\textnormal{ul}}^{\textnormal{H}}\boldsymbol{\textnormal{\textbf{A}}}_{\textnormal{T}}(\varPhi,\mathit{\Psi})-\boldsymbol{\textnormal{\textbf{H}}}_{\textnormal{dl}})(\boldsymbol{\textnormal{\textbf{G}}}\boldsymbol{\textnormal{\textbf{H}}}_{\textnormal{ul}}^{\textnormal{H}}\boldsymbol{\textnormal{\textbf{A}}}_{\textnormal{T}}(\varPhi,\mathit{\Psi})-\boldsymbol{\textnormal{\textbf{H}}}_{\textnormal{dl}})^{\textnormal{H}}\right\} .
\end{equation}

The $\textnormal{\textbf{G}}$ matrix is expressed as: $\textnormal{\textbf{G}}=((\boldsymbol{\textnormal{\textbf{H}}}_{\textnormal{ul}}^{\textnormal{H}}\boldsymbol{\textnormal{\textbf{A}}}_{\textnormal{T}}(\varPhi,\mathit{\Psi}))^{\textnormal{H}}\boldsymbol{\textnormal{\textbf{H}}}_{\textnormal{ul}}^{\textnormal{H}}\boldsymbol{\textnormal{\textbf{A}}}_{\textnormal{T}}(\varPhi,\mathit{\Psi})+\sigma^{2}\textnormal{\textbf{I}})^{-1}(\boldsymbol{\textnormal{\textbf{H}}}_{\textnormal{ul}}\boldsymbol{\textbf{H}}_{\textnormal{ul}}^{\textnormal{H}}\boldsymbol{\textnormal{\textbf{A}}}_{\textnormal{T}}(\varPhi,\mathit{\Psi})).$
Then, the final DL channel estimate is expressed as

\begin{equation}
\boldsymbol{\textnormal{\textbf{H}}}_{\textnormal{dl}}^{(2)}=\boldsymbol{\textnormal{\textbf{H}}}_{\textnormal{ul}}^{\textnormal{H}}\textnormal{\textbf{G\ensuremath{\boldsymbol{\textnormal{\textbf{A}}}_{\textnormal{T}}}(\ensuremath{\varPhi},\ensuremath{\mathit{\Psi}})}}.
\end{equation}

\section{Numerical Results }

We adopt an $8\times8$ URA transmit antenna setup at the BS and $2\times1$
receive antennas at the user side. The 3GPP FD spatial channel, as
described by (2)-(3), defines $C=12$ for line of sight (LoS) and
$C=20$ for non-LoS, with $Z=20$. The detailed simulation parameters
are listed in Table I. The performance of the proposed D-SCE is compared
with the state-of-the-art CSIT harvesting standards for FD-mMIMO channels
as follows:

\begin{table}
\caption{Simulation Parameters}
\centering{}%
\begin{tabular}{cc}
\hline 
Parameter & Value\tabularnewline
\hline 
Channel model & 3GPP-SCM {[}10{]}\tabularnewline
\hline 
Channel bandwidth & 10 MHz\tabularnewline
\hline 
BS antenna setup & 8 $\times$ 8 URA, $0.5\lambda$\tabularnewline
\hline 
UE antenna setup & 2 $\times$ 1 ULA, $0.5\lambda$\tabularnewline
\hline 
Spatial streams per UE & $d=1$ and $2$\tabularnewline
\hline 
Mobility condition & 30 km/hr\tabularnewline
\hline 
Azimuthal DoFs ($Q$) & $120$\tabularnewline
\hline 
Elevation DoFs ($U$) & $4$\tabularnewline
\hline 
\end{tabular}
\end{table}

$\vspace{0cm}$

\textbf{\textit{Beamformed CSI-RS}}

The original 3GPP proposal {[}8{]} has shown significant CSIT acquisition
gain. The enhanced-CSI-RS (E-CSI-RS) algorithm {[}9{]} is an extension
of the CSI-RS standard, proposed from \textit{Intel Research}. E-CSI-RS
algorithm adopts dual FD-DFT codebooks. The first $L-$codeword codebook
$\text{\textbf{\ensuremath{\mathcal{J}}}}$ defines the actual CSI-RS
span, physically beamformed across the FD cell space. Thus, the beamformed
DL channel $\boldsymbol{\textnormal{\textbf{H}}}_{\textnormal{dl}}^{bf}$
at the user side is de-beamformed to obtain an approximate estimate
of the actual channel $\boldsymbol{\textnormal{\textbf{H}}}_{\textnormal{dl}}^{approx}$
span given as 

\begin{equation}
\boldsymbol{\textnormal{\textbf{H}}}_{\textnormal{dl}}^{approx}=\boldsymbol{\textnormal{\textbf{H}}}_{\textnormal{dl}}^{bf}\mathcal{J}\text{\ensuremath{\mathcal{J}}}^{\textnormal{H}}.
\end{equation}

At the user side, the estimate of the DL full-span channel $\boldsymbol{\textnormal{\textbf{H}}}_{\textnormal{dl}}^{approx}$
is spatially projected over the second $N-$codeword codebook $\text{\textbf{\ensuremath{\mathfrak{T}}}}$.
Finally, each UE feeds-back its serving BS with an index $\varkappa$
of $\textnormal{B}=\log_{2}\left(N\right)$ bits, to select the closest-match
codeword to its estimated DL channel as 

\begin{equation}
\hat{\varkappa}=\arg\underset{\mathfrak{T}}{\max}||\boldsymbol{\textnormal{\textbf{H}}}_{\textnormal{dl}}^{approx}\text{\ensuremath{\mathfrak{T}}}||^{2}.
\end{equation}

Hence, the actual CSIT gain is $L$; however, the effectively harvested
gain is $N$, where $N\gg L$.

$\vspace{0cm}$

\textbf{\textit{Kronecker-based Beamforming}}

The KP-based CSIT approach {[}6{]} is an extension of the current
2D-MIMO CSIT standards. KP-based CSIT algorithms adopt the current
double-codebooks in LTE-Pro standards {[}5{]} for the azimuthal scanning,
where the azimuthal $t^{th}$ codeword, for $N_{h}=8$ horizontal
antenna setup, is given by 

\begin{equation}
\left(\mathcal{T}\right)_{t}=\left(\frac{1}{\sqrt{8}}\right)\left[\varpi,\,e^{\frac{j\pi t}{2}}\varpi\right]^{\textnormal{T}},
\end{equation}
where $\varpi=\left[1,e^{\frac{j2\pi t}{32}},e^{\frac{j4\pi t}{32}},e^{\frac{j6\pi t}{32}}\right]$.
For the elevation scanning, a DFT-based codebook $\textnormal{\textbf{a}}_{\kappa}^{v}(\varphi)$
is used, and the final FD-KP-based codebook is generated by $\boldsymbol{\textnormal{\textbf{W}}}(\theta,\varphi)=\text{\textbf{\ensuremath{\mathcal{T}}}}\,\otimes\textnormal{\textbf{a}}_{\kappa}^{v}(\varphi).$
\begin{figure}
\begin{centering}
\includegraphics[scale=0.52]{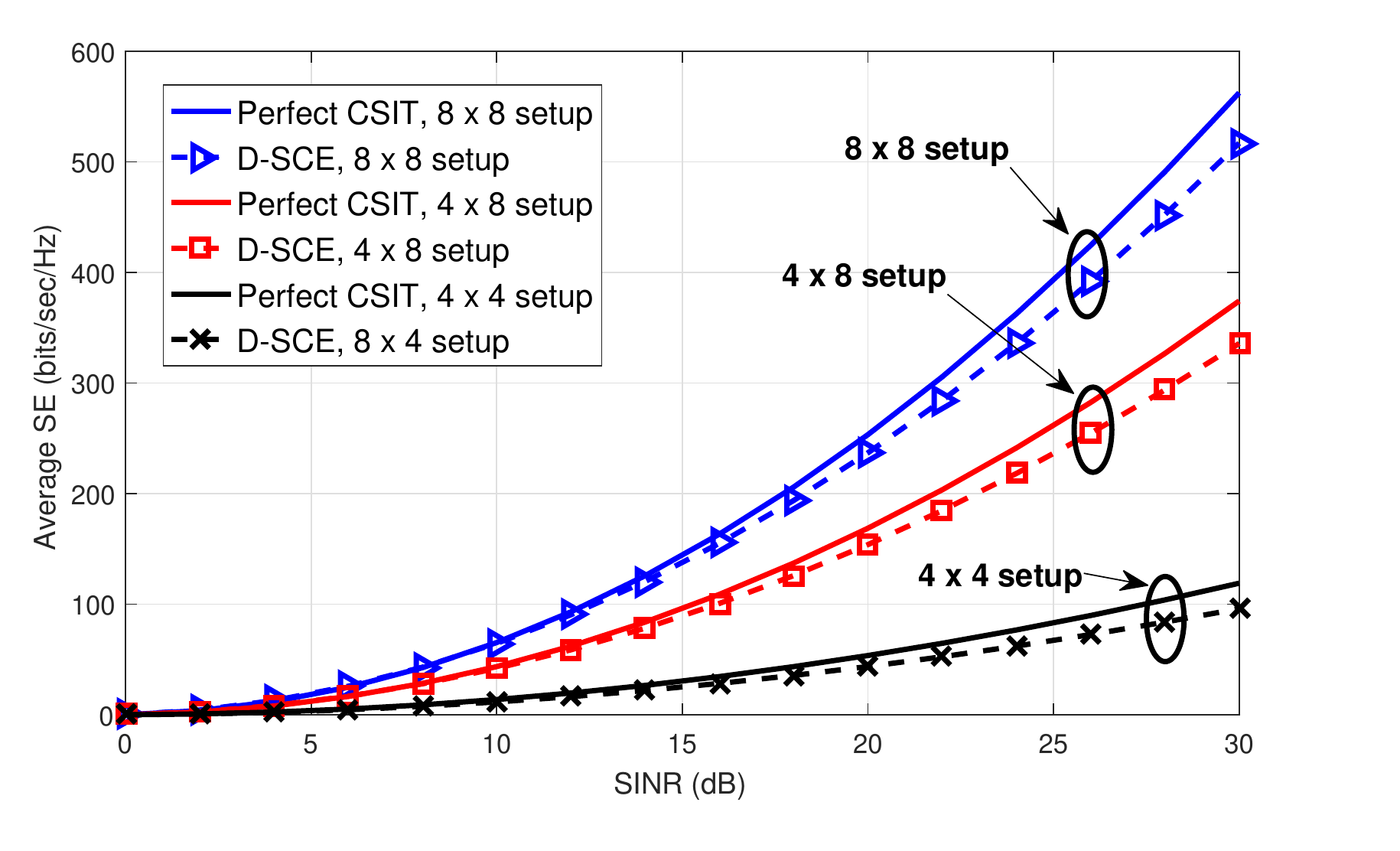}
\par\end{centering}
\centering{}\caption{Average SE performance of the perfect CSIT-based and the proposed
D-SCE algorithms, under different antenna setups, $d=2$. }
\end{figure}

Fig. 5 depicts the average SE performance in bits/sec/Hz of the perfect
CSIT-based and the proposed FD D-SCE algorithms, under different antenna
setups at the BS side, e.g., $8\times8$, $4\times8$ and $4\times4$
antenna arrays. The proposed D-SCE algorithm shows scalable SE with
the transmit antenna array size, approaching the perfect CSIT case
with $\textnormal{B}=0$ feedback overhead bits. However, to achieve
the perfect CSIT performance, the feedback overhead should be linearly
scaled with the size of the transmit antenna array as given by {[}2{]}
\begin{figure}
\begin{centering}
\includegraphics[scale=0.52]{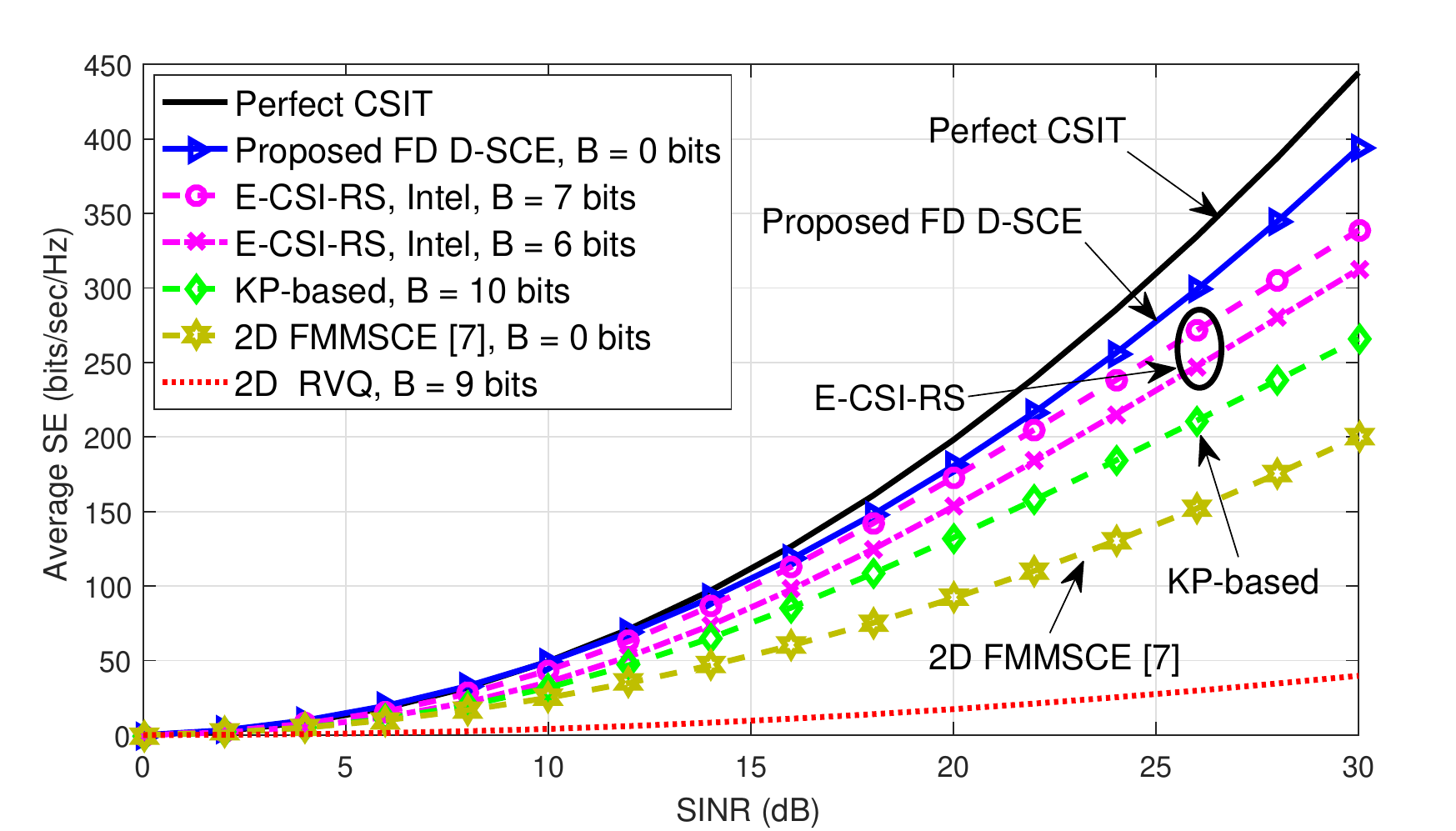}
\par\end{centering}
\centering{}\caption{{\small{}Average SE performance of the perfect CSIT, D-SCE, E-CSI-RS,
KP-based CSIT, 2D FMMSCE and 2D RVQ algorithms}, $d=1${\small{}.}}
\end{figure}

\begin{equation}
\textnormal{B}=(N_{t}-1)\log_{2}\textnormal{SNR},
\end{equation}
where SNR denotes the signal-to-noise-ratio, e.g., for $N_{t}=64$
and SNR$=10$ dB, the size of the feedback overhead is $\textnormal{B}=209.28$
per user per channel coherence time, which overwhelms the UL control
channel. 

~~~~Fig. 6 shows the SE performance comparison of the perfect
CSIT-based, proposed FD D-SCE, E-CSI-RS, KP-based, 2D FMMSCE {[}7{]}
and the 2D random vector quantization (RVQ) algorithms. The proposed
D-SCE with $\textnormal{B}=0$ feedback overhead bits outperforms
the E-CSI-RS with $\textnormal{B}=7$ bits per user per channel coherence
time, especially in the high SINR region, approaching the perfect
CSIT case. The E-CSI-RS suffers from performance degradation due to
the channel approximation, which contributes to the residual inter-user
interference. The proposed D-SCE provides significant outperformance
than the KP-based CSIT with $\textnormal{B}=10$ bits, e.g., 4-elevation
and 256-azimuthal codewords, respectively, due to the insufficient
CSIT precision of the adopted azimuthal LTE-Pro dual-codebook. FMMSCE
and conventional RVQ clearly suffer from performance degradation due
to the missing elevation beamforming, with $\textnormal{B}=0$ and
$9$ bits, respectively.

Moreover, the proposed D-SCE algorithm shows robustness to the UL
estimation error as: $\boldsymbol{\textnormal{\textbf{H}}}_{\textnormal{ul}}^{'}=\boldsymbol{\textnormal{\textbf{H}}}_{\textnormal{ul}}+\boldsymbol{\textnormal{\textbf{Y}}},$
where $\boldsymbol{\textnormal{\textbf{Y}}}\in C^{N_{t}\times M_{r}}$
is the Gaussian estimation error of the UL channel with variance $\sigma^{2}$.
As shown in Fig. 7, the D-SCE algorithm shows significant robustness
against the UL channel estimation error, when sufficient scanning
precision is available, e.g., maximum SE loss of 15 bps/Hz when $\sigma^{2}=0.3$,
with $Q=120$. This is due to the high density of the precise beamforming
directions with small spatial offsets around the optimal DL transmit
responses.  
\begin{figure}
\begin{centering}
\includegraphics[scale=0.35]{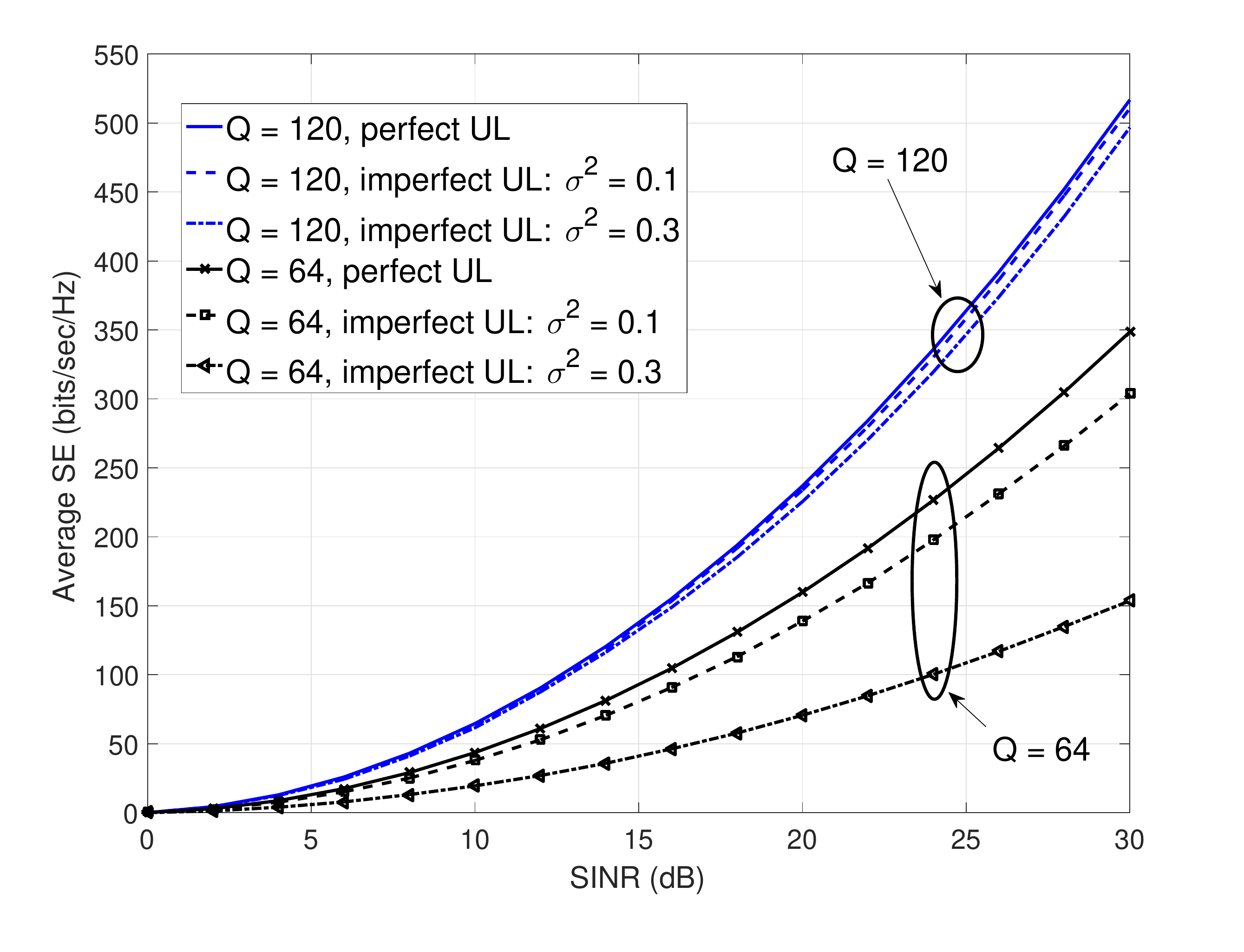}
\par\end{centering}
\centering{}\caption{{\small{}D-SCE robustness to UL channel estimation error. }}
\end{figure}

Furthermore, the precision of the azimuthal $Q$ and elevation $U$
DoFs is shown to influence the performance of the proposed D-SCE algorithm.
Fig. 8 and 9 present the mean loss of the overall SE with different
$Q$ and $U$ values. With low azimuthal scanning precision $Q$,
referenced to the $Q=120$ case,  the D-SCE algorithm exhibits consistent
performance degradation because the FD DoFs of the antenna array can
not be fully utilized, as shown in Fig. 8. Due to the small Doppler
shift in the elevation direction, and hence, the small spatial channel
elevation variance, selecting the elevation DoFs $U$ size does not
provide fast performance improvement. Hence, as depicted in Fig. 9,
referenced to the $U=4$ case used in this work, increasing the elevation
scanning precision $U\gg4$ does not offer significant spatial DoFs.
The performance gain of the proposed D-SCE algorithm is due to i)
the proper reduction of the spatial channel span, and ii) the \textit{on-the-go}
sufficient estimation of the principal transmit responses. 

\section{Conclusion}

A novel FD directional spatial channel estimation (D-SCE) algorithm
has been proposed. It blindly utilizes the statistical spatial correlation
between the UL and DL channels, to attain higher CSIT harvesting gain.
Compared to the state-of-the-art standard CSIT estimation algorithms,
the proposed D-SCE algorithm shows significant performance improvement
without FD CSIT overhead. With simple implementation complexity, zero
CSIT overhead, and scalability with the size of the transmit array,
the proposed D-SCE algorithm is a strong candidate for FD-mMIMO FDD
systems.
\begin{figure}
\begin{centering}
\includegraphics[scale=0.65]{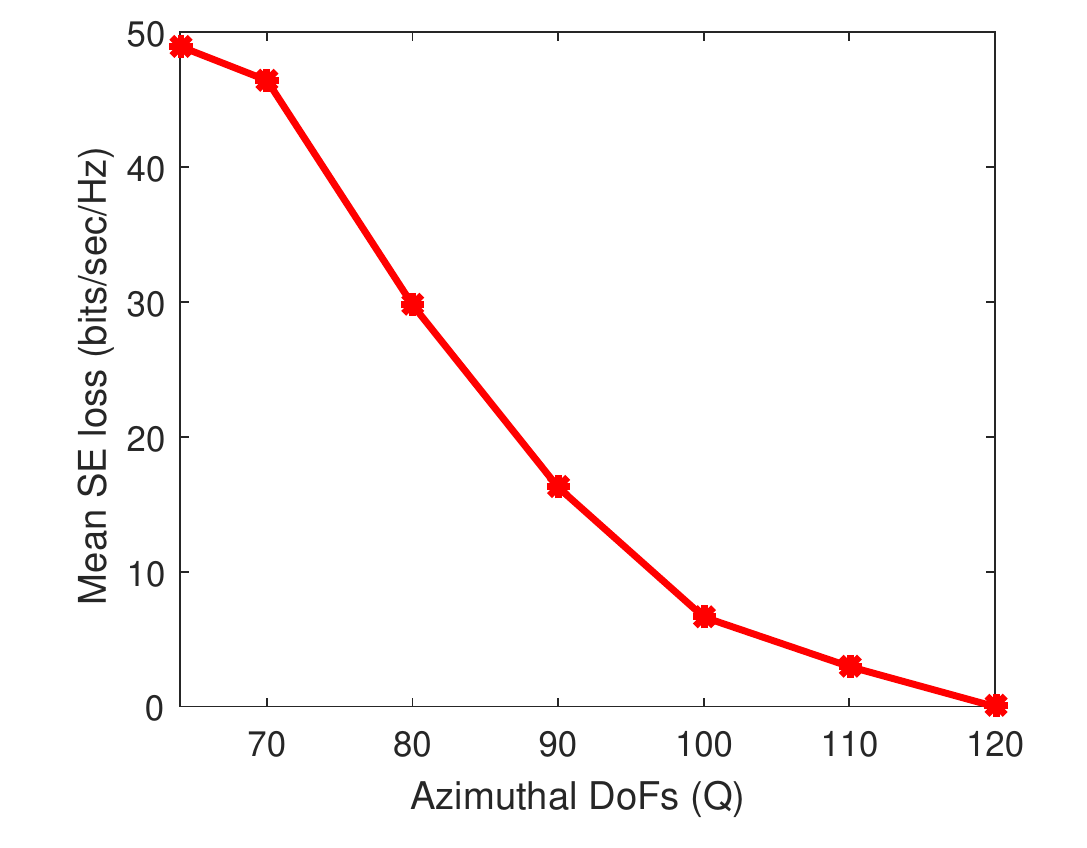}
\par\end{centering}
\centering{}\caption{{\small{}Selection of the azimuthal DoFs $Q$ size. }}
\end{figure}
 
\begin{figure}
\begin{centering}
\includegraphics[scale=0.65]{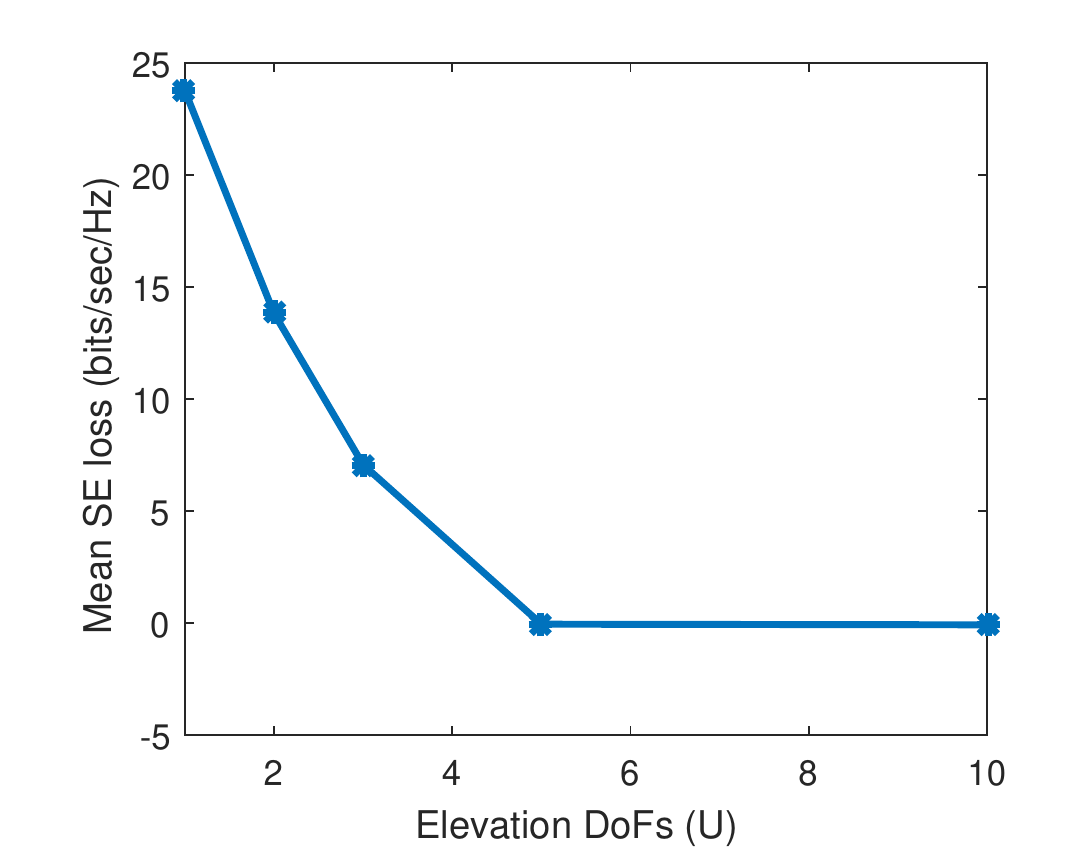}
\par\end{centering}
\centering{}\caption{{\small{}Selection of the elevation DoFs $U$ size. }}
\end{figure}

\end{document}